\documentclass[12pt]{nature}
\usepackage{amsmath,amsfonts,amssymb,amsthm,epsfig,array}
\usepackage{dsfont}
\usepackage{microtype} 
\usepackage[final]{pdfpages}
\usepackage{braket}
\usepackage{indentfirst}

\usepackage[hidelinks,colorlinks,linkcolor=blue,citecolor=blue,urlcolor=blue]{hyperref}
\usepackage[titletoc,title]{appendix}
\usepackage{times}
\makeatletter
\let\saved@includegraphics\includegraphics
\AtBeginDocument{\let\includegraphics\saved@includegraphics}
\renewenvironment*{figure}{\@float{figure}}{\end@float}
\makeatother

\def\bs#1{\boldsymbol{#1}}

\title{Anomalous circularly polarized light emission in organic light-emitting diodes caused by orbital-momentum locking}
\author{  Li Wan$^{1,2,3,\ast}$, Yizhou Liu$^{4}$, Matthew J. Fuchter$^{2,5}$, Binghai Yan$^{4,\dagger}$ \\
\normalsize{
$^1$ Department of Physics, Imperial College London, London, SW7 2AZ, United Kingdom\\
$^2$ Centre for Processable Electronics, Imperial College London, London, SW7 2AZ, United Kingdom\\
$^3$ Department of Physics, Chemistry and Biology (IFM), Link\"oping University, Link\"oping, 58431, Sweden\\
$^4$ Department of Condensed Matter Physics,Weizmann Institute of Science, Rehovot 76100, Israel\\
$^5$ Department of Chemistry, Molecular Sciences Research Hub, Imperial College London, White City Campus, London W12 OBZ, United Kingdom
}}

\begin{document} 

\maketitle 

\begin{abstract}

Chiral circularly polarized (CP) light is central to many photonic technologies, from optical communication of spin information to novel display and imaging technologies. As such, there has been significant effort in the development of chiral emissive materials that allow for the emission of strongly dissymmetric CP light from organic light-emitting diodes (OLEDs). A consensus for chiral emission in such devices is that the molecular chirality of the active layer determines the favored light handedness of CP emission, regardless of the light-emitting direction. Here, we discover that, unconventionally, oppositely propagating CP light exhibits opposite handedness and reversing the current-flow in OLEDs also switches the handedness of the emitted CP light. This direction-dependent CP emission boosts the net polarization rate by orders of magnitude by resolving an established issue in CP-OLEDs, where the CP light reflected by the back electrode typically erodes the measured dissymmetry. Through detailed theoretical analysis, we assign this anomalous CP emission to a ubiquitous topological electronic property in chiral materials, namely the orbital-momentum locking. Our work paves the way to design new chiroptoelectronic devices and probes the close connections between chiral materials, topological electrons, and CP light in the quantum regime.\end{abstract}

\section*{Introduction}
Chirality characterizes parity symmetry-breaking where a molecule cannot be superposed on its mirror image in chemistry and biology\cite{Siegel1998,barron2021symmetry}. Chiral enantiomers exhibit opposite chiroptical activity when coupling to light\cite{Kuball2017,barron1986symmetry}. In physics, chirality usually refers to the spin-momentum locking of particles such as Weyl fermions\cite{Armitage2017,Yan2017} and the circular polarized light. Chiral organics were recently reported to exhibit a topological feature \cite{liu2021chirality}, in which the electronic orbital and momentum are locked together, to rationalize the intriguing spin selectivity in DNA-type molecules\cite{Naaman2019,evers2021theory}.
Hence given the intimate relationship between electronic states and light-matter interactions, we were inspired to raise a question: Can topological electronic properties  {(i.e., orbital-momentum locking)} enhance chiroptical activity and therefore advance the rapidly developing (chir)optoelectronic technology\cite{brandt2017added,Shang2022}? 

A future industrial application of organic chiral emissive materials is in circularly polarized organic light-emitting diodes (CP-OLEDs)\cite{zhang2020recent}, which should eliminate the $\sim$50\% internal light loss caused by the contrast-enhancing circular polarizer in OLED displays. Such efficiency gains occur via direct circularly polarized electroluminescence (CP-EL) from the CP-OLED, which can pass through the contrast-enhancing polarizer unhindered.\cite{yang2013induction} The effectiveness of this strategy depends on the degree of circular polarization of EL, where higher polarization gives better efficiency for the display in the presence of such polarizers.\cite{Wan2020}

Since the first CP-OLED reported in 1977\cite{doi:10.1021/ja971912c}, the  {circularly-polarized electroluminescence} of  {a} material was also assumed to be identical to  {circular polarization} measured  {in absorption and in photoluminescence (cases without current flow) from the same electronic transition} . In another words, CP-EL was considered nearly the same process as CP photoluminescence (CP-PL) [or the inverse process of optical circular dichroism (CD)]  due to a shared electronic transition, and the magnitude of CP emission determined by the product of electric and magnetic transition dipole moments.\cite{craig1984molecular,Greenfield2021} Thus most efforts in this field were made on developing more twisted chiral emitters with stronger magnetic transition dipoles to improve optical chirality\cite{yan2019configurationally,adfm202010281}, without taking current flow in an OLED device into consideration.

More importantly, in terms of device engineering, the reflective back-electrode in an OLED device is another key issue. In all prior studies of chiral emissive materials, CP emission is conventionally expected to exhibit the same handedness in both emission directions (forward and back) from the point of recombination, thus any back reflection within the device will invert the handedness of CP emission travelling backwards and cancels out the forward CP emission, reducing the net EL circular polarization that exits the device through the transparent electrode\cite{yan2019configurationally,zinna2017design,GrellAM,Zinna2022}.
Consequently, the magnitude of EL circular polarization from devices is much smaller than the corresponding CP-PL  {measured in transmittance geometry,} which does not suffer issues of reflection\cite{zinna2017design}(Fig. \ref{fig:1}a). Even though constructing semi-transparent OLEDs can, to some extent, mitigate the problem of reflection, such a strategy reduces the overall the device performance in a displays, negating the original intention of energy saving at the polarizer.\cite{yan2019configurationally}

Among all CP-OLEDs reported  {and many other chiral optoelectronic devices based on 2D\cite{doi:10.1126/science.1251329} and perovskite materials \cite{doi:10.1126/science.abf5291}}, chiral polymeric materials\cite{Wan2020,wan2019inverting,wan2021strongly,Lee2017a,DiNuzzo2017} demonstrate significant circular polarization in PL and EL, several orders of magnitudes stronger than other chiral emissive systems\cite{li2020axially,yan2019configurationally,Li2021anie,Li2018anie} (see Fig. \ref{fig:2}a). Despite the analysis above, when constructing optoelectronic devices from such materials, their CP-EL remains equal, or sometimes is even enhanced compared to CP-PL or CD. Although previous theoretical\cite{laidlaw2021factors,swathi2020supramolecular} and experimental\cite{Wan2020,wan2019inverting,wan2021strongly} work attributed the strong optical circular dichroism to a predominately excitonic origin, these analyses cannot account for  the comparable or enhanced circular polarization in EL devices, given the expected detrimental effect of back-electrode reflection. 

In this work, we discover an anomalous light emission phenomenon from chiral polymeric CP-OLEDs. For the chiral polymeric materials under study, CP-EL exhibits opposite handedness in forward and backward emission directions, counter-intuitive to what is usually expected in EL or PL (Fig. \ref{fig:1}b). With such direction-dependent CP emission, the back-reflected light exhibits the same handedness as the forward emission, avoiding the polarization cancellation which occurs in devices using other materials and boosting the net CP-EL exiting the device\cite{yan2019configurationally,zinna2017design}. 
Furthermore, for the first time, we explain the effect of current flow on CP-EL, where its handedness can also be switched by reversing the current flow in an OLED. We propose that the directional CP-EL observed is caused by the topological nature of the electronic wave functions in chiral polymers. 
Because of orbital-momentum locking\cite{liu2021chirality}, \textcolor{black}{
the current flow induces nonequilibrium orbital polarization in electron and hole carriers. 
Therefore, finite angular momentum transfers from electron/hole orbital to the photon spin in the optical transition. When they have the same spin, the counter-propagating CP lights exhibit opposite handedness. }
This orbital polarization effect rationalizes the fact that the handedness of CP light is determined both by the current direction and the emission direction. 
 {Furthermore, this model reveals an exotic CP-EL mechanism caused by current-induced time-reversal breaking.}
Our work paves they way to design novel chiroptoelectronic devices with strong circular polarization.

\section*{Results}

\begin{figure}
    \centering
    \includegraphics[width=\textwidth]{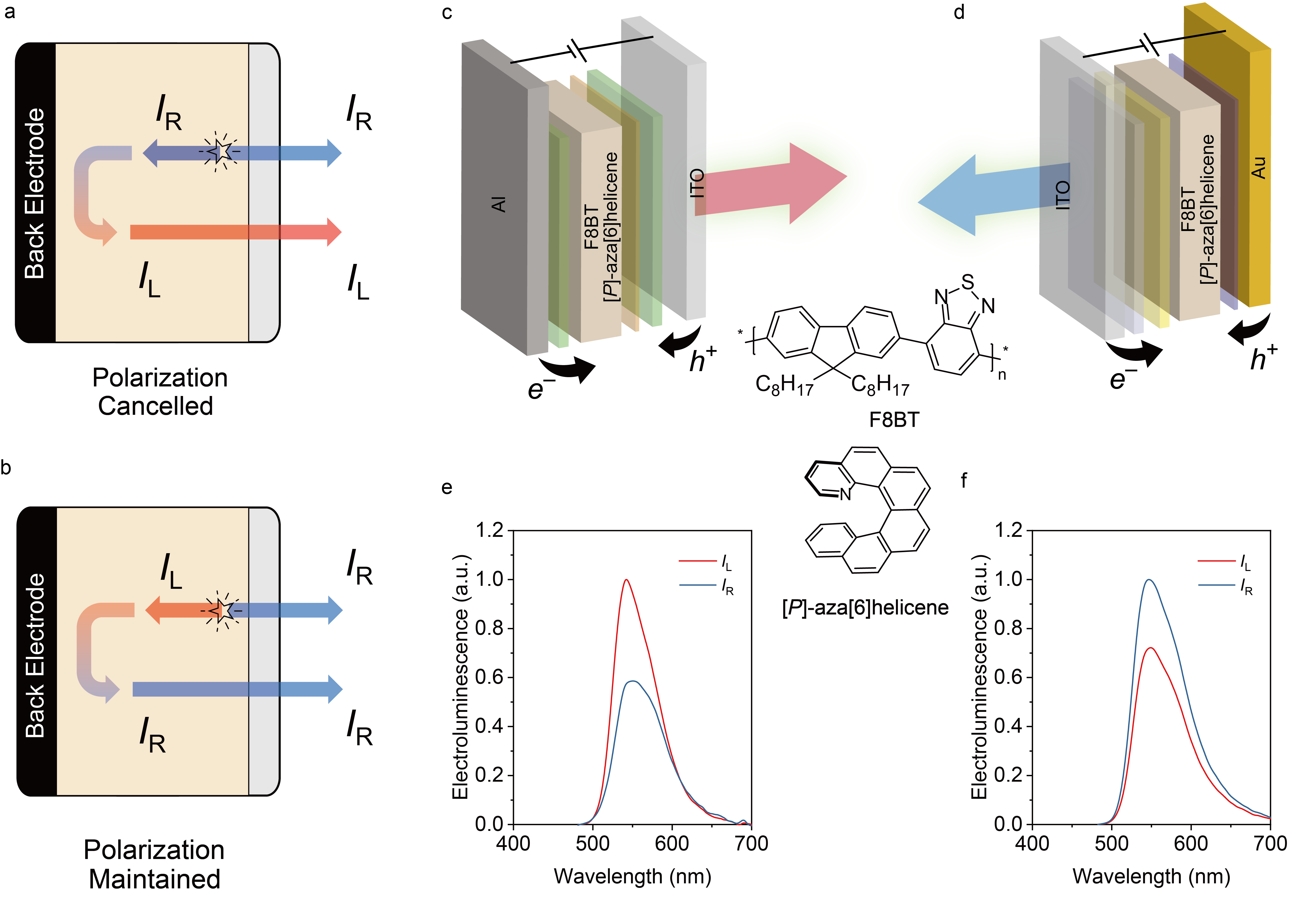}
    \caption{\textbf{Device-structure dependent circular polarized (CP) Electroluminescence (EL) .} (a) Normal circular polarization effect where CP-EL is emission direction-independent. (b) Anomalous circular polarization effect where CP-EL is emission direction-dependent. $I_{L}$ and $I_{R}$ represent light intensity of Left- (red arrows) and Right-handed (blue arrows) circularly polarised emission. Device structures of (c) conventional and (d) inverted CP-OLEDs. EL of F8BT:[\textit{P}]-aza[6] helicene-based CP-OLEDs recorded from (e) conventional and (f) inverted CP-OLEDs. Inset: Molecular structure of F8BT and [\textit{P}]-aza[6]helicene.}
    \label{fig:1}
\end{figure}

A chiral polymer blend consisting of an achiral light-emitting polymer (i.e., poly(9,9-dioctylfluorene-\textit{alt}-benzothiadiazole), F8BT, Fig. \ref{fig:1}) and a non-emissive chiral additive (i.e., [\textit{P}]-aza[6]helicene) was selected for the investigation of CP-EL.  {Upon thermal annealing of spin-casted thin-films, the chiral additive (10 wt\%) induces a} strong and robust  {chiral structure and} optical CD  {to the originally achiral polymer with an} absorption dissymmetry factor ($g_{abs}$) of $\sim 0.6$ (see Fig. S1)\cite{yang2013induction,wan2019inverting}, calculated in the following way:
\begin{equation}
    g_{abs} = \frac{A_L - A_R}{A} = \frac{\Delta A}{A},
\end{equation}
where $L/R$ stands for left/right-handed CP light and $A$ refers to the absorbance. To investigate how emission direction affects the CP-EL, CP-OLEDs were fabricated using both conventional and inverted device structures where the transparent electrode (i.e., Indium Tin Oxide, ITO) serves as anode and cathode, respectively (Figs. \ref{fig:1}c and \ref{fig:1}d).

\begin{figure}
    \centering
    \includegraphics[width=\textwidth]{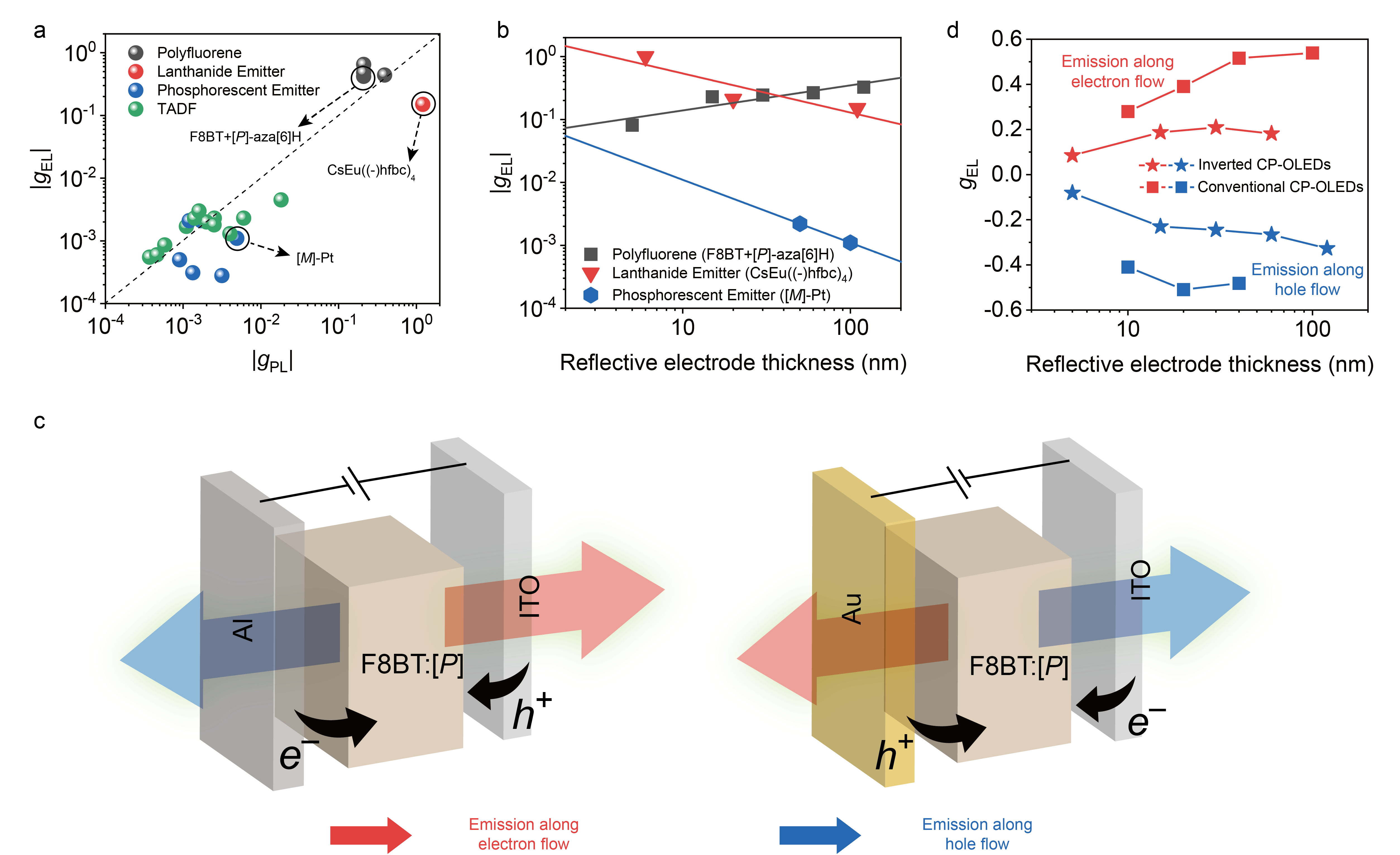}
    \caption{\textbf{Emission-direction-dependent CP emissions relative to charge carrier flow directions. }
    (a) Summary of reported CP-OLEDs using different types of chiral emitters including polyfluorene, lanthanide complex, thermally assisted delayed fluorescent emitter (TADF), and phosphorescent emitters. Detailed information on molecular structures and corresponding citations can be found in Table S1 and Scheme S1. $|g_{PL}|$ and $|g_{EL}|$ represent photo- and electro-luminescence dissymmetry factors. (b) Comparison of $|g_{EL}|$ values reported from polyfluorene (inverted device in this work), lanthanide complex (data points from Ref.\citeonline{zinna2017design}), and phosphorescent emitter (data points from Ref.\citeonline{yan2019configurationally}) as a function of reflective electrode thickness. The trends are represented as the solid lines. (c) Schematic diagram of semi-transparent conventional (top) and inverted (bottom) CP-OLEDs based on F8BT:[P]-aza[6]helicene. Red arrows stand for the EL emitted along the electron flow direction and blue arrows stand for the EL emitted along the hole flow direction. Interlayers are removed for clarity. (d) Summary of $g_{EL}$ measured from both sides of the semi-transparent CP-OLEDs. Detailed spectra for data point in b and d can be found in Figs. S3 and S4.}
    \label{fig:2}
\end{figure}

With a fixed current flow direction (Figs. \ref{fig:1}c and \ref{fig:1}d), when measuring the EL through transparent ITO, left-handed CP light is observed in a conventional device where the light emits in the same direction as the electron injection. A positive EL dissymmetry factor $g_{EL}$ of +0.54 can be calculated following: 
\begin{equation}
    g_{EL} = \frac{I_L - I_R}{(I_L + I_R)/2},
\end{equation}
where $I_{L/R}$ is the irradiance recorded from the CP-OLEDs. However, despite a fixed absolute stereochemistry of chiral material in the emissive layer of both devices, the sign of the CP-EL signals was found to be dependent on the device structure. When the emission direction relative to the current direction is switched, the inverted CP-OLED emits right-handed circularly polarized light through ITO with a $g_{EL}$ of $?0.33$. 
Apart from the emission direction-dependent CP-EL signals in conventional versus inverted devices, we detected no evidence of erosion of $g_{EL}$ by the reflective electrodes. Compared with other reported CP-OLEDs\cite{li2020axially,yan2019configurationally,Li2021anie,Li2018anie}, the polyfluorene-based CP-OLEDs we developed exhibit one of the highest known $g_{EL}$ values (Fig.~\ref{fig:2}a). In contrast, lanthanide complexes exhibit intrinsically high PL dissymmetry ($g_{PL}$)\cite{zinna2017design}, but the $g_{EL}$ recorded from the transparent electrode of lanthanide-based CP-OLEDs dramatically decreases when increasing the thickness of the reflective metal electrode. This is similarly observed in other small molecule CP-OLEDs\cite{li2020axially,yan2019configurationally,Li2021anie,Li2018anie} (Fig.~\ref{fig:2}b).

To compare our results with other previously reported CP-OLEDs, we performed CP-EL measurements on semi-transparent OLEDs in both conventional and inverted CP-OLEDs (Fig.~\ref{fig:2}c). Surprisingly, emission direction-dependent CP-EL behavior was observed in both device structures, where the CP-EL from forward and backward emission (i.e., through a semi-reflective electrode) exhibit the opposite handedness. Considering this emission direction-dependent dissymmetry factor is only observable in EL but not for CP-PL or CD of the chiral thin films (Fig. S2), we speculate that this behavior is associated with the flow of charge carriers within the devices. To unambiguously describe and compare the emission direction-dependent CP-EL signals in two device architectures, we define the emission direction relative to the charge carrier flow direction (See Fig.~\ref{fig:2}c). Specifically, emission from the transparent ITO in the conventional device and the emission from semi-transparent Au in the inverted device are defined as directionally aligned with the electron flow. Conversely, emission from the opposite electrodes in conventional and inverted devices is defined as directionally aligned with the hole flow. In contrast to other chiral emitters in OLED devices\cite{yan2019configurationally,zinna2017design}, we find $|g_{EL}|$ of net emission from the transparent electrode increases when increasing the thickness of the reflective electrode (Figs.~\ref{fig:2}b and ~\ref{fig:2}d) for our chiral polymeric materials. As summarized in Fig.~\ref{fig:2}d, regardless of device structure used, F8BT:[\textit{P}]-aza[6]helicene based CP-OLEDs demonstrate positive $g_{EL}$ when emission occurs along the direction of electron flow and switch to negative $g_{EL}$ when emission occurs along the hole flow direction. This indicates that the previously observed inversion of CP-EL sign from conventional versus inverted devices (see Fig.~\ref{fig:1}) is due to emission direction-dependent CP-EL relative to the electron flow direction. Given this outcome, the reflective electrode in our system results in reflected light with an identical sign to the forward emission, resulting in a comparable or boosted net CP EL signal (Fig.~\ref{fig:2}b).

\section*{Discussions}

\subsection{ {Experimental Summary}}
Our CP-EL data exhibits profound features, distinct from those obtained with other chiral emitters in OLEDs (Fig. 2b). Besides extremely large $g_{EL}$, the CP-EL handedness can be switched both by the current flow direction and by the light emission direction.  {By carefully examining the optical elements in our blend thin films\cite{wade2020natural}, the sign inversion observed here is not caused by linear (LD/LB) effects nor circularly polarized scattering. There are negligible morphological change when the film is casted on the PEDOT:PSS or ZnO layer, which suggests that the active layer morphology is identical between conventional and inverted devices (Fig. S5). Previously, for CP-OLEDs constructed from other materials, the position of the recombination has been reported to effect the magnitude of the CP-EL\cite{wan2019inverting,Lee2017a}. By using a TPBi electron transport layer (LUMO $\sim$2.7 eV) in our conventional devices, we generate a larger electron injection barrier from Ca ($\sim$2.9 eV) than would be the case when directly depositing Ca onto F8BT (LUMO $\sim$3.3 eV). We observe TPBi emission in our EL spectra, which indicates the recombination zone in our conventional devices occurs at the F8BT/TPBi interface, away from the ITO electrode. Similarly, in our inverted OLED, TCTA was used to shift the recombination to the F8BT/TCTA interface till the observation of the interface recombination (Fig. S6)\cite{doi:10.1021/acs.jpcc.7b11039}. Therefore, by careful selection of contact and transport layers, the recombination zones in both devices were pinned physically at the far interface of F8BT, away from ITO electrode. \textcolor{black}{This ensures the CPEL sign inversion investigated here is not caused by the previously reported active layer thickness/recombination zone\cite{wan2019inverting,Lee2017a}, but instead by current-direction (See sign-unchanged data in Fig. S7)}.}

 {By ruling out other origins, we demonstrate that the CP-EL handedness inversion is sensitive to both the current flow direction and the light emission direction.}
In other words, the emitted photon seems to retain 'Circularly polarised light emitted by OLEDs exhibits opposite handedness depending on the propagation direction of the light. Switching the current flow in the OLED also switches the light handedness'a memory of the propagating direction of recombining electrons and holes, despite the fact that initial carrier velocities are usually ignored to understand the light emission process\cite{https://doi.org/10.1002/pi.1974}. Such an outcome indicates salient  {current-flow-induced
time-reversal symmetry (TRS)-breaking in the system. The exciton-coupling model\cite{laidlaw2021factors,swathi2020supramolecular} is based on quantum states with TRS and cannot explain such current-direction- and emission-direction-dependent CP-EL.}

\subsection{ {Theoretical model}}
It is known that helicene additive promotes the F8BT polymer to form chiral assemblies in solid-state thin films\cite{laidlaw2021factors,wade2020natural}. The chiral polymer blend is the real-space channel for both current flow and light emission.  {In the following, we will discuss the light emission in the presence of TRS-breaking due to the current flow.}


 {We first will revisit the general theory that describes the CP emission effect and interpret our experiments from an anomalous term.} According to Fermi's golden rule, the emission rate of CP light is, 
\begin{equation} \label{eq:Fermi}
    I_{L/R}=\frac{2\pi}{\hbar} |\bra{0}H^\prime \ket{1}|^2 \delta(\epsilon_1 - \epsilon_0 - \hbar \omega)
\end{equation}
where $\ket{0}$ ($\ket{1}$) represents the ground (excited) state with energy $\epsilon_0$ ($\epsilon_1$),  $H^\prime$ is the light interaction Hamiltonian $H^\prime= -e \textbf{E} \cdot \textbf{r} - \textbf{m} \cdot \textbf{B}$ with $\textbf{m}$ being the magnetic moment, $\textbf{E}$ and $\textbf{B}$ are the light electric and magnetic fields, respectively, and $\hbar \omega$ is the photon energy. 
For right/left-handed light traveling along the $z$ axis, the electric and magnetic fields are $\textbf{E}=E_0 (1,\pm i,0)/\sqrt{2}$ and $\textbf{B}=E_0/c(\mp i,1,0)/\sqrt{2}$.  Therefore, the leading term of CP light emission can be derived as,
\begin{equation}\label{eq:EL}
    I_L - I_R = -\frac{2\pi e^2}{\hbar} I_0~ \mathrm{Im}[x^{01}y^{10}]\delta  \\
    - \frac{2\pi e}{\hbar c} I_0~ \mathrm{Im} [m_{x}^{01} x^{10} + m_{y}^{01} y^{10}]\delta
\end{equation}
where $x^{01}=\bra{0}x\ket{1}$ and $m_{x}^{01}=\bra{0}m_{x}\ket{1}$ represent the electric and magnetic transition dipoles, respectively, and $I_0=|E_0|^2$. We note $\textbf{m}=(m_x,m_y,m_z)$, $\textbf{r}=(x,y,z)$ and $\delta$ for the same $\delta$-function in Eq. \ref{eq:Fermi}.

The second term in Eq.~\ref{eq:EL} is routinely employed to understand CD, CP-PL or CP-EL for organic/inorganic systems and has been called natural chiroptical activity\cite{riehl1999chiroptical}. 
Because it corresponds to interaction with the electric and magnetic fields of light, this term involves both electric and magnetic transition dipoles. We call it the normal circular polarization effect (NCPE) in the following. In chiral organic molecules, the OAM change in the transition is zero because TRS forces the net OAM of $\ket{0,1}$ to be zero. Due to the conservation of angular momentum, the total spin angular momenta of all emitted photons sum to zero. Because counter-propagating photons emitted are equal in probability, they carry opposite spins, i.e., they share the same handedness (see illustration by Fig.~\ref{fig:3}c). Thus, the handedness of CP emission is free from the emission direction in the case of NCPE, as observed in PL and EL reported in other chiral emissive systems\cite{Gaspar2015,li2020axially,yan2019configurationally,Li2021anie,Li2018anie}.


The first term in Eq.~\ref{eq:EL} corresponds to the dipole interaction with light electric field. If $\ket{0,1}$ respect TRS, this term is exactly zero. Additionally, a similar term corresponds to the Berry-curvature-induced\cite{Xiao2010} anomalous Hall effect\cite{Nagaosa2010} in solid crystals and also vanishes if TRS appears.  Therefore, the first term was generally ignored when studying organic molecules. It was referred to as the magnetic CD~\cite{Souza2008,Yao2008} in absorption of magnetic materials or in a external magnetic field. However, if electrons and holes carry finite velocities before recombination, the first term cannot be naively neglected. In other words, the current flow can induce magnetization, more specifically the orbital magnetization as we will show. The nonequilibrium phase breaks TRS in chiral molecules. 
We refer to the first term as the anomalous circular polarization effect (ACPE) here. In such a case, ACPE may contribute more to the net circular polarization than NCPE because the electric field is much stronger than magnetic field in light.

\subsection{ {Orbital-momentum locking in chiral molecules}}
 {We discuss the TRS-breaking of $\ket{0,1}$ in the presence of current flow. Although the average momentum is zero 
for electrons in a molecule, the simultaneous momentum ($k$) can be significant. For example, we can approximate the molecule as a box confining electrons inside (see Figs.~\ref{fig:3}a-b ). Electrons move back-and-forth due to boundary scattering and form a standing wave $\ket{\psi}$. The standing wave is a superposition of a forward-moving state $\ket{\psi^\pm}$ and a backward-moving state $\ket{\psi^-}$}, where we choose $\ket{\psi^+} = \ket{\psi^-}^*$ because TRS allows $\ket{\psi}$ to be real valued. If an electron picks up a velocity along $\pm z$ caused by the current flow, its wave function reduces from $\ket{\psi}$ to $\ket{\psi^\pm}$. We point out that $\ket{\psi^\pm}$ themselves violate TRS, although $\ket{\psi}$ does not. 

Next, $\ket{\psi^\pm}$ carry opposite OAM ($\pm l$) in a chiral molecule~\cite{liu2021chirality}.  {As illustrated in Fig.~\ref{fig:3}b, electrons that travel along a chiral pathway pick up a self-rotation, i.e., the OAM, in analogue to the spinning bullet out of a rifled barrel. Mathematically,} we can generally describe a positive-moving plane wave by $\ket{\psi^+}=A(\rho,z)e^{il\phi+ikz}$, where $A(\rho,z)$ is a general coefficient depending on $z$ and the radial distance $\rho$, $k$ is the momentum, and $l=0,\pm 1, \pm 2?$ represents OAM.
$\ket{\psi^+}$ is a chiral plane wave if $l \neq 0$ and reduces to a normal plane wave for $l=0$. Because both the inversion symmetry and mirror symmetry, either of which forces $l=0$, are broken  $l \neq 0$ generically holds in a chiral system. It is obvious that $\ket{\psi^-}$ with $-k$ carries opposite OAM $-l$ because $\ket{\psi^+}=\ket{\psi^-}^*$. Such orbital($l$)-momentum($k$) locking represents the wave function topology, in which the parallel or antiparallel $l-k$ relation depends on the molecular chirality and chemical potential. { It is similar to the monopole-like spin-momentum locking in the topological Weyl fermion~\cite{Armitage2017,Yan2017}. }

\subsection{ {Angular momentum transfer from electrons to light}}
 {
The injected electrons and holes carry finite OAM because of the polymer chirality in the OLED. Given the low mobility in the organic semiconductor, the linear momentum is quickly relaxed, for instance, by interface scattering between neighboring aggregate clusters. But the OAM relaxation time should be much longer than the momentum relaxation time because neighboring clusters share the same chirality. Due to the same chirality protection, we also expect that OAM is robust against electron-electron interactions. Therefore, electrons and holes can preserve the OAM polarization when they form excitons for light emission. In the following, we discuss the OAM transfer from carriers to CPL in the electron-hole recombination. }

\begin{figure}
    \centering
    \includegraphics[width=\textwidth]{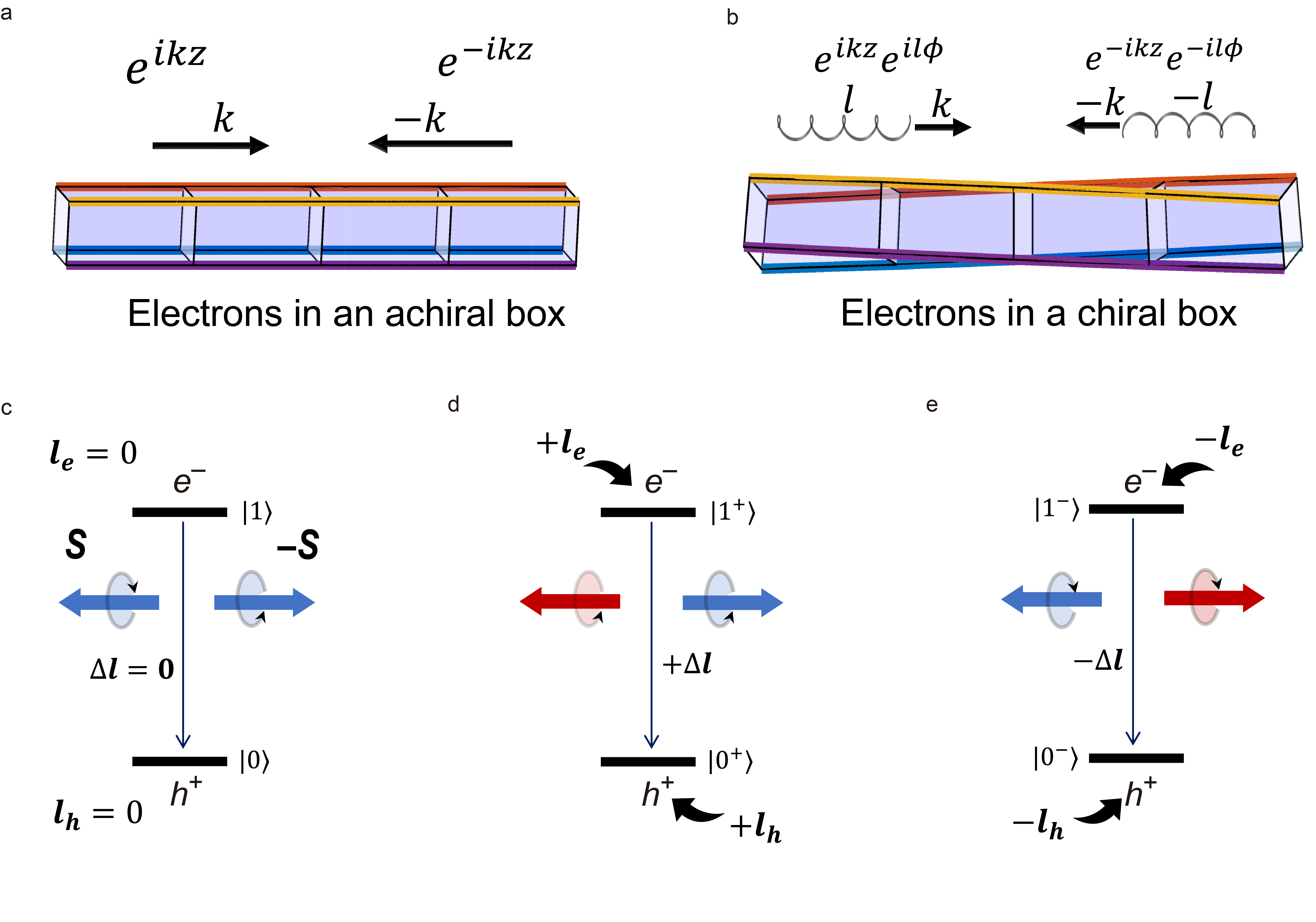}
    \caption{Illustration of orbital-momentum locking and anomalous circular polarization effect.  {(a) The wave function ($\ket{\psi}$) of The electron confined in an achiral box can be regarded a superposition of counter-propagating plane waves,$\ket{\psi^{\pm}}\sim e^{\pm ikz}$.
    (b) The wave function in a chiral box can be regarded a superposition of counter-propagating chiral plane waves, $\ket{\psi^{\pm}}\sim e^{\pm ikz} e^{\pm l\phi}$.  $\ket{\psi^{\pm}}$ carry opposite orbital angular momenta (OAM) ($\pm l$), i.e. $\ket{\psi^{\pm}}$ share the same \textit{electronic chirality} induced by the box chirality. The black arrow represents the propagating direction ($k$) and the helix-like trajectory indicates $l$.} (c) Normal circular polarized emission from the excited state $\ket{1}$ to ground state $\ket{0}$, where $\ket{0,1}$ carry zero OAM. The angular momentum transfer in the transition $\Delta l =0$. So the zero net spin of light leads to the same handedness in emitted lights along two directions. ``e'' and ``h'' represent electron and hole, respectively. (d) Anomalous circular polarized emission from $\ket{1^+}$ to $\ket{0^+}$, where $\ket{0^+,1^+}$ carry finite OAM. Because $\Delta l$ is finite, oppositely emitted lights carry the same spin and exhibit opposite handedness.  {(e) Reversing the current-flow with respect to (d) switches the CPL handedness.}
    }
    \label{fig:3}
\end{figure}

The ACPE term in Eq.~\ref{eq:EL} is equivalent to the OAM shift in the optical transition\cite{Oppeneer1998,Souza2008,Yao2008} \textcolor{black}{
\begin{equation}
 \Delta l = \braket{\hat{L}_z}_{0\rightarrow1}=x^{01}p_y^{10}-y^{01}p_x^{10},   
\end{equation}}
 where $\hat{L}_z = xp_y - yp_x$ is the OAM operator (see more information in Supplementary Information).
In the presence of current along $-z$, we need replace $\ket{0^+}$ ($\ket{1^+}$) for $\ket{0}$ ($\ket{1}$) to evaluate the ACPE in Eq.~\ref{eq:EL}. In this case, $\Delta l$ is nonzero in the optical transition from $\ket{1^+}$ to $\ket{0^+}$, both of which carry finite OAM, as illustrated in Fig. \ref{fig:3}d. In addition, reversing current leads to $-\Delta l$ (Fig. \ref{fig:3}e). We note that $\Delta l$ is gauge invariant although $l$ of a given band depends on the specific gauge.

For a photon with a positive spin angular momentum , it is defined as right-handed CP light if it travels along $+z$ and as the left-handed CP light if along $-z$. Therefore, the finite angular momentum , which is transferred  from OAM of electrons/holes to spin of photons ($S = +\Delta l$), forces that the handedness of CP light relies on the emission direction. Overall, the orbital-momentum locking eventually leads to the current direction- and emission-direction dependencies of CP-EL, fully consistent with our experiments. 

 {To summarize the ACPE mechanism, it includes two parts. (i) In a chiral molecule/chiral aggregate, the OAM is parallelly or anti-parallelly locked to the intrinsic momentum (see more discussions on its difference from the mobility-related drift velocity in SI). Opposite momenta with opposite OAM coexist and cancel each other at equilibrium. (ii) When the current flows, one momentum is enhanced while the opposite one is suppressed, so as the OAM. Therefore, conducting electrons/holes are OAM polarized in the nonequilibrium phase. Direct electron-hole recombination leads to finite OAM transfer from charge carriers to CPL, leading to the ACPE. }

\subsection{ {Ab initio calculations}}
\begin{figure}
    \centering
    \includegraphics[width=\textwidth]{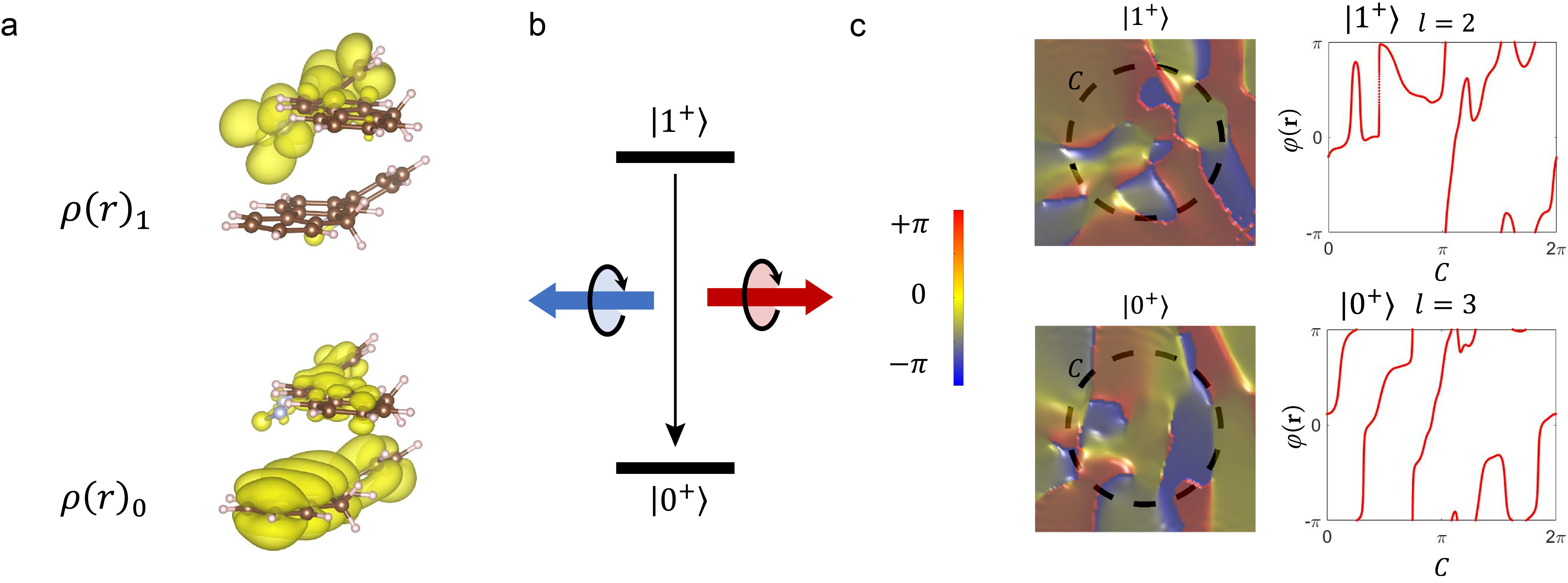}
    \caption{The chirality and circular polarized emission for stacking two F8BT molecules with a twisted angle. (a) The total charge density distribution for the HOMO ($\ket{0}$) and LUMO ($\ket{1}$) states calculated by density functional theory. (b) Schematics of transition from $\ket{1^+}$ to $\ket{0^+}$. 
     {
    (c) The phase distribution $\varphi(\mathbf{r})=l\phi$ in the $xy$ plane for a vortex-like plane wave $e^{ i(l\phi \pm k z)}$ (left). The $l$ is deduced from the phase winding number of $\varphi(\mathbf{r})$ (right) counterclockwise along the dashed circle path $C$. $\ket{0^+}$ ($\ket{1^+}$) with $+k$ exhibits $l= 3$ ($ 2$) while $\ket{0^-}$ ($\ket{1^-}$) with $-k$ exhibits $l= -3$ ($ -2$), demonstrating the orbital-momentum locking ($l||k$).}
    }
    \label{fig:4}
\end{figure}

Furthermore, we quantitatively estimate the ACPE and NCPE for the chiral F8BT polymer assembly by \textit{ab initio} calculations. It is challenging to refine the accurate atomic structure of such chiral aggregates. Without losing generality, we simulate chiral stacking of F8BT molecules and focus on the intermolecular chirality that is associated with the dominant charge transport direction along the layered packing structure (noted as $z$ axis here)\cite{donley2005effects}. 
Although $\ket{0,1}$ can be generally many-body wave functions, we use the highest occupied molecule orbital (HOMO) and lowest unoccupied molecule orbital (LUMO) to represent $\ket{0}$ and $\ket{1}$, respectively, by ignoring higher-order corrections (like the distortion in excited states) in the calculations. As shown in Fig. \ref{fig:4}, two layer stacking with a counter-clockwise twisting angle 30$^{\circ}$ reshapes HOMO and LUMO wave functions dramatically compared to a single layer of molecule. By calculating the ACPE involving the $+z$ moving HOMO and LUMO, we obtain a large dissymmetry factor
$|g_{EL}| =0.48$ ($0.44$) for two (three) layer stacking, which is in the same order of magnitude as experimental $g_{EL}$. The OAM can be evaluated from the phase winding number in the $xy$ plane  {(see Fig.~\ref{fig:4}c and the Methods section)}, verifying the orbital-momentum locking in $\ket{0^\pm}$ and $\ket{1^\pm}$. Because $\ket{0,1}$ is usually composed by many plane waves, the total value of $l$ is unnecessarily an integer. Better knowledge on the molecular arrangement of chiral polymer assemblies will help improve the prediction power of calculations in the future work. 

Additionally, the current-induced magnetization in our experiments is relevant to the orbital rather than spin of electronic states. If electron-spin polarization matters, it would require substantial spin-orbit coupling (SOC) in the device. We know that these organic polymers made of light elements exhibit negligible SOC. Despite that metal electrodes may include heavy elements, the circular polarization rate remains the same for Al, Ag and Au electrodes with largely varied SOC (Fig. S8). Thus, the role of electron spin may be negligible in ACPE although ACPE may also appear in systems with strong SOC. The ACPE is caused by the chirality-induced orbital polarization, different from the chirality-induced spin selectivity effect discussed in literature\cite{Naaman2019}. Our experiment demonstrates a significant consequence of the electron orbital effect in chiral materials. Here, the electron OAM is indicated by the the handedness flip of CP light (spin flip of light) as reversing the current flow (e.g., see Figs.~\ref{fig:1}c-d).

\section*{Summary}
In summary, we report an anomalous phenomenon where the handedness of CP light emission depends on the emission direction. This effect enables us to design unconventional CP-OLED devices with large $g_{EL}$ and without errosion from the back-electrode reflection. 
We highlight that the orbit-momentum locking causing ACPE is strongly associated with the charge transport mode in the polymer systems and therefore suggest the following design principles for further development of CP-OLEDs with strong CP-EL.  {To ensure the entire stack of molecular assemblies exhibit strong ACPE, it is necessary that the emissive sites should strongly couple with chiral transport sites or ideally within the same sites as in our polymer systems, to induce strong orbital-momentum locking in conduction electrons. } If charge carriers are independently transported, such as in host materials, then they get scattered to random adjacent chiral emissive sites, the net momentum and OAM will be quenched and only NCPE will appear. In this case, CP-EL can no longer be considered as the same origin as CP-PL and circular dichroism where no charge transport and current flows exist.

We propose an ACPE that involves finite angular momentum transfer in the optical transition. Because ACPE and NCPE come from the first and second-order optical transitions in Eq.~\ref{eq:EL}, ACPE is often much larger than NCPE when TRS is broken. We highlight that the unusual TRS-breaking in ACPE is driven by the nonequilibrium orbital magnetization, which originates in the chiral orbital nature in wave functions.  {In CP-OLED, such orbital magnetization is caused by the current flow (rather than static magnetization or magnetic field), the impact of which was rarely recognized in previous studies on chiral materials\cite{Barron2007lord}. Our work reveals an intriguing unification of chirality in seemingly unrelated aspects: structure geometry, electronic topology, and the light handedness.
The chirality information can be transferred from the material geometry to electronic wave function and further to the spin of light. 
}

\newpage
\noindent \textbf{Acknowledgements:} 
 Dedication: To the memory of Professor Alasdair James Campbell (1961-2021). L.W. would like to thank Prof. Alasdair Campbell and Prof. Jenny Nelson for useful discussions at Imperial College London. L.W. would like to thank Mr. Kim Stjerne (Biolab A/S, Denmark) and Dr. Theis Brock-Nannestad (University of Copenhagen, Denmark) for the access to CPL-300 spectrometer. L.W. would like to thank Dr. Rui Zhang (Link\"oping University) for the help of processing GIWAXS data. B.Y. acknowledges the financial support by the MINERVA Stiftung with the funds from the BMBF of the Federal Republic of Germany and the European Research Council (ERC Consolidator Grant ``NonlinearTopo'', No. 815869). L.W. and M.J.F. would like to thank Cambridge Display Technology Limited (company number 02672530) for providing the F8BT polymers. L.W. and M.J.F. would like to acknowledge EPSRC research grants EP/P000525/1, EP/L016702/1 and EP/R00188X/1. For the purpose of open access, the author has applied a Creative Commons Attribution (CC BY) licence to any Author Accepted Manuscript version arising
\\
\noindent \textbf{Competing Interests:} The authors declare the following competing financial interest(s): M.J.F. is an inventor on a patent concerning chiral blend materials (WO2014016611). The remaining authors declare no competing interests.\\

\newpage
\section*{References}

\section*{Methods}
\textbf{Samples:} [\textit{P}]-aza[6]helicene was synthesized as previously reported\cite{wan2019inverting} and separated using preparative chiral HPLC.

\textbf{Thin-film and device fabrication:} The cleaning process for all substrates (fused silica and prepatterned ITO glass, Thin Film Devices Inc., 20 ohms/sq) involved rinsing in an ultrasonic bath with acetone, isopropyl alcohol (IPA), Hellmanex III (Sigma-Aldrich), and deionized water for 30 min. These were transferred to a plasma asher for 3 min at 80 W and 50 W before spin-coating for fused silica and prepatterned ITO, respectively. F8BT and aza[6]helicene were dissolved in toluene to a concentration of 30 mg/mL and blended to form a 10\% aza[6]helicene solution. About 130 nm thick emissive layer can be achieved by dynamically spin-coating at 2300 rpms for 1 min. Chiral samples were annealed for 10 min in a nitrogen atmosphere (glovebox, $<0.1$ ppm of H$_2$O, $<0.1$ ppm of O$_2$). Dynamic coating ensures strong chiroptical activity is achieved without giving too thick films, compared to previous studies\cite{wan2019inverting,Wan2020}. Organic film thicknesses was monitored using a Dektak 150 surface profiler and the metal thickness was used as displayed from QCM monitor. 

 {\textbf{Photophysical characterization:}} Circular dichroism measurements were performed using a Chirascan (Applied Photophysics) spectrophotometer.  {CP-PL was measured in a transmittance geometry using a CPL-300 JASCO spectrometer. GIWAXS data was measured in ALBA beamline (Spain). The incidence angles of the X-ray beam were set to be 0.2 for
all films. The GIWAXS patterns were recorded with a 2D CCD detector and an X-ray irradiation time 10-20s, dependent on the saturation level of the detector.}

\textbf{CP-EL:} Left-handed and right-handed CP emission spectra were collected using a combination of linear polarizer and zero-order quarter-wave plate (546 nm, Thorlabs) placed before detectors. The device is driven by a constant current of 1 mA (pixel area = 0.045 cm$^2$). The dissymmetry factor $g$ in the EL spectra was calculated from the equation $g = 2(I_L ? I_R)/(I_L + I_R), |g| \le 2$. Here, $I_L$ and $I_R$ are the left-handed and right-handed emission intensities. EL spectra from the PLED were recorded using an Ocean Optics USB 2000 charge-coupled spectrophotometer. All CP-EL measurements are carried out after measuring with only a linear polarizer to ensure negligible linear polarization or random polarization, which should be considered in all CP emission measurements. 

\textbf{Conventional devices:} Poly(3,4-ethylenedioxythiophene):poly(styrenesulfonate) (PEDOT:PSS) (H.C. Starck GmbH) (55 nm) was dynamically spin-coated on cleaned ITO substrate with a spin speed of 2500 rpms for 1 min. [\textit{P}]-aza[6]helicene-blended F8BT deposition was the same as for the thin film studies, followed by the thermal evaporation of a 10 nm TPBi (Sigma-Aldrich), 10 nm Ca layer (Sigma-Aldrich), capped by an Al (99.99\%, Kurt J. Lesker Company Ltd.) layer up to 100 nm on the organic layer under a vacuum level of $1 \times 10^{?7}$ mbar. Conventional device structure: ITO/PEDOT:PSS/TFB/ F8BT:[\textit{P}]-aza[6]helicen/TPBi/Ca/Al.

\textbf{Inverted devices:} The ZnO is deposited on the cleaned ITO by a sol?gel method which has been described elsewhere followed by a PEIE (30 wt \% in water, Sigma-Aldrich) rinsing step. Then, [\textit{P}]-aza[6]helicene-blended F8BT was dynamically spin-coated onto ZnO/PEIE. Afterward, 25 nm TCTA (97\% Sigma-Aldrich), MoOx (99.97\%, Sigma-Aldrich) and Au (99.99\%, Kurt J. Lesker Company Ltd.) up to 120 nm were thermally evaporated onto the organic layer under vacuum level of $1 \times 10^{?7}$ mbar.Inverted device structure: ITO/ZnO/PEIE/ F8BT:[\textit{P}]-aza[6]helicene/TCTA/MoOx/Au. 

\textbf{Ab inito calculations:} The electronic wave functions and charge densities of F8BT chiral assemblies were calculated by the density-functional theory (DFT) implemented in the Vienna \textit{ab initio} Simulation Package (VASP).\cite{Kresse1996PRB} The generalized gradient approximation (GGA) was used for the exchange-correlation functionals.\cite{perdew1996generalized} 
F8BT molecules were stacked along the $z$ axis in a twisted manner. The molecular cluster model included at least 10 $\AA$ vacuum distances along all directions.  An energy cutoff of 400 eV was used for the plane wave basis. The total number of plane waves is about 4$\times10^5$. We note the plane wave by wave vector $\mathbf{G}=(G_x, G_y, G_z)$. In the plane wave basis, for example, the wave function of HOMO, $\ket{0}$, can be expressed as:
\begin{equation}
\begin{split}
    \ket{0} &= \sum_{G_z} \phi_{0}(G_z) e^{iG_zz}, \\
    \ket{0^\pm} &= \sum_{G_z>0 / G_z <0} \phi_{0}(G_z) e^{iG_zz},
\end{split}
\end{equation}
where $\phi_{0}(G_z) = \sum_{G_x, G_y} c^{0}_{\mathbf{G}} e^{i(G_x x + G_y y)}$ and $c^{0}_{\mathbf{G}}$ is the plane wave coefficient extracted directly from DFT wave functions. The charge density is $\rho_{0} = |\ket{0}|^2$. Besides $G_z z$, the phase of the $G_z$ propagating wave in Fig.~\ref{fig:4} is arg$\phi_{0}(G_z)$. 
Under electrical current along $-z$, the CP electric dipole transition amplitudes are calculated as:
\begin{equation}\label{I_RL}
    \begin{split}
        I_{R/L}(\omega) &\propto \left|\langle 0^+ | x \pm iy | 1^+ \rangle\right|^2 \delta(\epsilon_1 - \epsilon_0 - \hbar\omega) \\
        &= \left|\int d\bs{r}~ \sum_{G'_z>0} \sum_{G_z>0} \phi^*_0(G'_z)(x \pm iy) \phi_1(G_z) e^{i(G_z-G'_z)z}\right|^2 \delta(\varepsilon_1 - \varepsilon_0 - \hbar\omega) \\
        &= \left|\sum_{G_x} \sum_{G_y} \sum_{G_z>0} c^1_{\mathbf{G}} \left( \frac{\partial c^0_{\mathbf{G}}}{\partial G_x} \mp i\frac{\partial c^0_{\mathbf{G}}}{\partial G_y} \right)^*\right|^2 \delta(\varepsilon_1 - \varepsilon_0 - \hbar\omega),
    \end{split}
\end{equation}
where $^*$ represents taking complex conjugate. 
One can analyze the symmetry constrain on CP emission from Eq.~\eqref{I_RL}. If inversion symmetry is present, then $c^{0,1}_{\mathbf{G}} = p_{0,1}c^{0,1}_{-\mathbf{G}}$ with $p_{0,1}=\pm 1$ refers to the parity eigenvalue of $\ket{0}$ or $\ket{1}$. On the other hand the time-reversal symmetry requires that $c^{0,1}_{\mathbf{G}} = (c^{0,1}_{-\mathbf{G}})^*$. Thus, if inversion and time-reversal symmetries exist simultaneously, the coefficient $c^{0,1}_{\mathbf{G}}$ is purely real (imaginary) for the parity even (odd) state. Therefore, $I_R=I_L$ always holds according to Eq. \eqref{I_RL}.  When inversion is broken, the circular polarization ($I_R \neq I_L$) can appear.

 {\textbf{OAM and winding number:} The OAM $l$ is a topological number charactering the phase winding of the wave function in real space, which is defined as \cite{Nakahara2017book}:
\begin{equation}
    l = \frac{1}{2\pi} \oint_C d\varphi = \frac{1}{2\pi} \oint_C d\bs{r} \cdot \bs{\nabla_r} \varphi(\bs{r}),
\end{equation}
where $\varphi(\bs{r})$ is the real-space distribution of wave function phase, and $C$ refers to the integration contour (dashed circle in Fig. \ref{fig:4}c). $l$ represents the total phase change (in multiples of $2\pi$) after circulating the contour $C$. Generally, a nonzero winding number indicates a topological defect inside the contour $C$ which cannot be removed by continuously varying the wave function without creating or annihilating another topological defect.
}

\section*{Data and code availability:} All data needed to evaluate the conclusions in the paper are present in the paper and/or the Supplementary Materials. Additional data related to this paper can be requested from the authors. Source data are provided. \\
\noindent {$^\ast$Email of correspondence:li.wan@liu.se}\\ 
\noindent {$^\dagger$ Email of correspondence: binghai.yan@weizmann.ac.il}

\section*{Method-only references:}


\end{document}